\documentclass[14pt,a4paper,twoside,fleqn]{extarticle}
\usepackage{els-chem-crc}
\usepackage{times}

\usepackage{graphicx}
\usepackage[figuresright]{rotating}
\configuresideways
\usepackage{amssymb}
\usepackage{amsfonts}
\usepackage{amsmath}

\chapter{xxxx}
\title{ Effective models for   charge transport in 
DNA nanowires }

\author{
        R. Gutierrez\address{Institut f\"ur Theoretische Physik, 
	Universit\"at Regensburg, D-93040 Regensburg, Germany}
        and
        G. Cuniberti
 }

\begin{document}

\maketitle

\begin{abstract}
     The rapid progress  in the field of molecular electronics has led to an
      increasing  interest on DNA oligomers as
       possible components of electronic circuits at the nanoscale. For this,
        however, an understanding of 
   charge transfer and transport mechanisms in this
 molecule is required.  Experiments show that a large number of factors
 may influence the electronic properties of DNA. Though full
 first principle approaches are the ideal tool for a theoretical 
 characterization of
 the structural and electronic properties of DNA, 
the structural complexity of this molecule
 make these methods of limited use. Consequently, model Hamiltonian
 approaches, which filter out single factors influencing charge
propagation in the double helix are highly valuable. 
In this chapter, we
  give a review of
 different DNA models which are thought  to capture the 
   influence  of some of
 these factors.  We will specifically focus on static and dynamic disorder. \\ \ \\	     
\end{abstract}

Keywords: DNA conduction, static disorder, electron-vibron interaction, correlated disorder, dissipation

\maketitle

\section{Introduction}

The increasing demands on the integration densities of electronic devices 
are  considerably
limiting conventional semiconductor-based electronics. 
As a result, new 
possibilities have been explored in the last decade. They have led to the 
emergence of molecular electronics, which  basically relies on the idea 
of using 
single molecules or molecular groups as   elements of electronic
devices. A new conceptual 
idea advanced by molecular electronics is the switch from a
top-bottom approach, where the devices are extracted from a single 
 large-scale building block, to a bottom-up approach in which the 
whole system is composed of small basic building blocks with recognition
and self-assembly properties.

A molecule that has recently  attracted the attention of both, experimentalists and
theoreticians, is DNA. The observation of electron transfer between 
intercalated donor 
and acceptor centers in DNA oligomers in solution
 over unexpected long length scales~\cite{murphy93},
 led to a revival of interest in the conduction properties of this molecule. 
 Though the idea that DNA might be  conducting  
is rather old~\cite{spivey62}, there were no conclusive proofs that it 
 could support charge
transfer over long distances.  This is however a critical issue 
when considering e.~g.~damage repair during the 
replication process~\cite{fried03}. Apart from the  
 relevance  of these and similar experiments for biology and genetics, 
 they also suggested 
that by appropriately tuning the experimental conditions, 
DNA molecules might be able 
to carry an electrical current. Further, DNA oligomers might be useful 
as templates in molecular electronic circuits, by exploiting their 
self-assembling and self-recognition properties~\cite{DR,keren03,pompe99}. 
Though many technical and theoretical 
problems have still to be  surmounted,
it is possible nowadays to carry 
out transport experiments on single molecules connected to metallic 
electrodes. 

However, despite the many expectations put on DNA as a potential ingredient of 
molecular electronic  circuits, transport experiments 
on this molecule have revealed a very intriguing and partly 
contradictory behavior. 
Thus, it has been found that DNA may be 
insulating~\cite{braun98,storm01}, semiconducting ~\cite{porath00,cohen05} or
 metallic~\cite{yoo01,tao04}. 
These results demonstrate the high sensibility of DNA transport  to 
different factors affecting charge motion, like 
the quality of the contacts to the metal electrodes, the base-pair sequence, 
the charge injection  
into the molecule or  environmental effects (dry vs. 
aqueous environments)  among others.

Theoretically,  knowledge of the electronic structure 
of the different building units of a DNA molecule (base pairs, sugar and 
phosphate groups) is essential
for clarifying  the most effective transport mechanisms. First principle approaches are the most suitable 
tools for this goal.
However, the huge complexity of DNA makes {\it ab initio}
calculations still very demanding, so that only comparatively 
few investigations have
 been performed
 ~\cite{felice02,barnett01,gervasio02,artacho03,soler03,lewis97,star04,sankey04,mehrez05}.
Further, environmental effects such as the presence of hydration shells
 and counterions  make {\it ab initio} calculations even more 
 challenging~\cite{barnett01,gervasio02,endres05}.
 
 In this chapter, we will review a complementary (to first principle approaches)
 way to look at DNA, namely, 
 model Hamiltonians. They play a significant role  in filtering out 
 possible charge transfer and transport mechanisms as well as in guiding the
 more involved 
 first principle  investigations. We are not aiming at a thorough 
 review of Hamiltonian-based theories. In fact, since the authors belong to 
 the ``physical 
 community'', model approaches for charge transfer formulated in the ``chemical 
 community'' will not be the scope of this chapter. The interested reader can 
 consult  e.g.~Refs.~\cite{schuster04,nitzan01,nitzan04,jortner98,jortner02,berlin01}. We are also 
not considering the influence of electron-electron interactions onto charge  transport, an issue that needs further clarification~\cite{yi03,apalkov05}
 In the next two sections, we discuss models describing the influence of 
 static disorder and dynamical effects,  on charge propagation in
 DNA. For the sake of the presentation, we discuss both factors in different
 sections. Nevertheless, the reader should be aware that an interplay between
 them is expected  to be closer to  reality.

\begin{figure}[t]
\centerline{
\includegraphics[width=6.2in,height=3.0in]{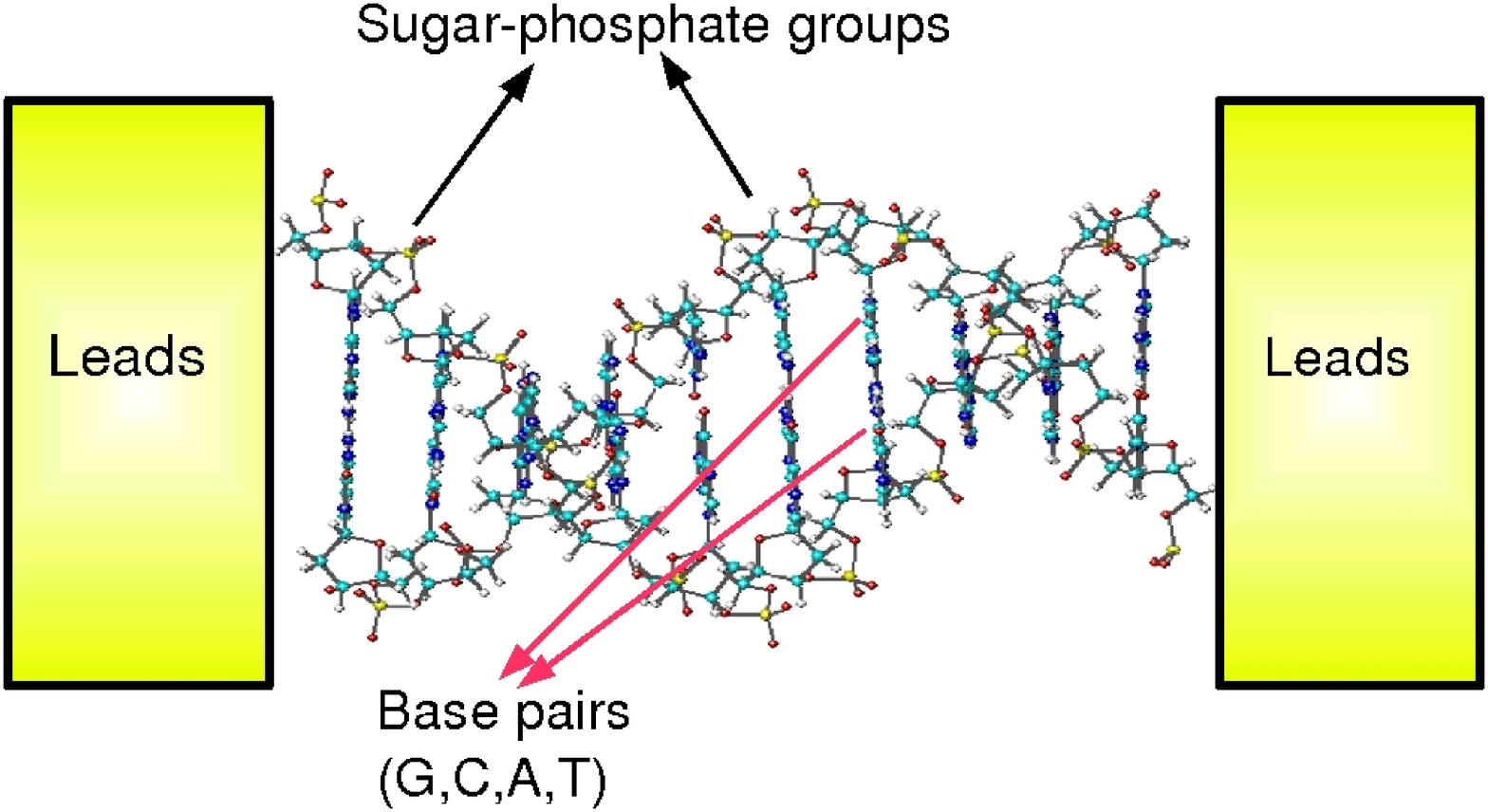}%
}
\caption{\label{fig:fig1}%
Schematic representation of a double-stranded DNA oligomer with an
arbitrary base-pair sequence and connected to left and right electrodes.}
\end{figure}

\section{Static Disorder}
DNA oligomers consist of four building blocks (oligonucleotides): 
adenyne (A), tymine (T), cytosine (C) and guanine (G). As is well-known, they 
have specific binding properties, i.~e.~only A-T and G-C pairs are possible, see Fig.~1. 
Sugar and phosphate groups ensure the mechanical stability of the double helix and
protect the base pairs. Since the phosphate groups are negatively charged, 
the topology of the duplex is only conserved if it is immersed into an aqueous 
solution containing counterions (Na$^{+}$, Mg$^{+}$) that neutralize the 
phosphate groups. Thus, experiments on ``dry'' DNA usually means that the 
humidity has been strongly reduced, but there are still water molecules 
and counterions attached to the sugar-phosphate mantle.

 The specific base-pair sequence is obviously essential for DNA to fullfil
 its function as 
a carrier of the genetic code. However, this same fact can be detrimental 
for charge transport. The 
apparent random way in which  the DNA sequence is composed strongly 
suggests that a charge propagating along the 
double helix may basically feel a random potential leading to backscattering. 
It is well-known that in a one-dimensional system 
with uncorrelated disorder all electronic states are completely  localised
(Anderson localization). However, 
correlated disorder with e.~g.~power-law correlations~\cite{carpena02} may 
lead to delocalised states within some special energy windows in the 
thermodynamic limit, the exact structure of the spectrum being determined by the so called scaling exponent $\alpha$.
This quantity  describes 
the correlation properties of a random process~\cite{peng92,carpena02},
specifically, the length-dependence of the position autocorrelation function:
$C(l)\sim l^{-\alpha}$. 
Thus,  $\alpha=0.5$ corresponds to a pure random walk, while other values
indicate the presence of long-range correlations and hence, the absence 
of relevant length scales in the problem (self-similarity).

Some of the main issues to be addressed when investigating the role of 
disorder in DNA  are, in our view, the following:
 (i) Is the specific base-pair  sequence in DNA completely random (Anderson-like) 
 or do there exist (long- or short-ranged) correlations?  
(ii) A measure for 
the degree of confinement of the electronic wave function  is given 
by the localization length $\xi$~\cite{philips_book}. 
 Are the resulting  localization lengths larger or smaller than the 
actual length $L$ of the DNA segments studied in transport 
experiments? 
For $\xi\gg L$ the system may appear as effectively 
conducting, despite the presence of disorder, though in the 
thermodynamic limit all states may remain localised. To clear these issues 
requires a close cooperation between experimentalists and theoreticians. 
In what follows we review some theoretical studies addressing these problems.

The simplest way to mimic a DNA wire is by assuming 
that after charge injection, the electron (hole) will basically propagate 
along one of the strands (the inter-strand coupling being much smaller), 
so that one-dimensional tight-binding chains can be a good 
starting point to minimally describe a DNA wire. 
Roche~\cite{roche03} investigated such a model for Poly(GC) and $\lambda$-phage
DNA,  with on-site disorder (resulting 
from the differences in the ionization potentials of the base pairs)
and bond disorder $\sim\cos{\theta_{n,n+1}}$  related to twisting motion 
of nearest-neighbor bases along the strand, $\theta_{n,n+1}$ 
being independent Gaussian-distributed random variables. 
 Poly(GC)  displays two electronic bands, thermal fluctuations reduce the 
transmission peaks and also slightly, the band widths. The effect of disorder 
does not appear to be very dramatic. In the case of 
 $\lambda$-phage, however, 
the transmission peaks are considerably diminished 
in intensity and in number with increasing chain length at zero temperature, 
since only few electronic states are not backscattered by the 
random potential profile of the chain. Interestingly, the average Ljapunov 
exponent, which is related to the localization length, increases
with increasing temperature, indicating that despite thermal fluctuations
 many states are still contributing to  charge transport.

In an early paper  Roche {\em et al.}~\cite{roche03a} used 
 scaling coefficients (Hurst exponents), which usually indicate  the existence 
of  long-range correlations in disordered systems. 
Their results show that e.~g.~DNA built from Fibonacci sequences has a
 very small Hurst exponent (indicating strong correlations). 
Uncorrelated random sequences show a strong fragmentation
and suppression of the transmission with increasing length, while in 
correlated sequences several states appear to be rather 
robust against the increasing rate of backscattering. Hence, it may be expected 
that correlated disorder will be more favorable for long-distance carrier
transport in  DNA wires.

Another typical example of correlated 
disorder was presented by Alburquerque {\em et al.}~\cite{albur05} 
within a one-dimensional tight-binding model. The authors investigated 
the  quasi-periodic 
Rudin-Shapiro sequence as well as the 
human genome Ch22.  
As expected, the transmission bands became more
 and more fragmented with increasing number 
of nucleotides. 
Though for very long chain lengths all electronic states did tend to be 
completely localised, long-range correlations
yielded large localization lengths and thus transport might still be 
supported for special energy points on rather long  wires.

Zhu {\em et al.}~\cite{guo04} formulated an effective tight binding model
 including only HOMO and 
LUMO of poly(GC) together with onsite Coulomb 
interactions.  Onsite and off-diagonal disorder, related to fluctuations
 of the local electrostatic potential~\cite{adessi03} 
 and to twisting motion of the base pairs at 
finite temperatures, respectively, were also included.
 The main effect of the Coulomb interaction was 
 to first reduce the band gap, so that the system goes over 
to a metallic state, but finally the gap reappears as 
a Coulomb-blockade gap. Twisting disorder
 was apparently less relevant for short wires and low temperatures.

 A very detailed study of the localization 
properties of  electronic states in two  minimal 
models of different DNA oligomers (poly(GC), $\lambda$-DNA, telomeric DNA)
 was presented by D. Klotsa {\em et al.}~\cite{klotsa05}: 
a fishbone model~\cite{gio02,gmc05a,gmc05b} 
 and a ladder model. 
 Both models fullfil the minimal requirement of showing a band-gap in the 
 electronic spectrum, mirroring the existence of a HOMO-LUMO gap in isolated 
 DNA molecules.  
 However, the ladder model 
  allows for an inclusion of interstrand effects as well as to include 
the specific base-complementarity typical of the DNA duplex, an issue 
that can not be fully captured by the first model. 
The authors were mainly interested in environmental-induced disorder. 
Hence, they assume that only the backbone sites 
were affected by it, while the nucleotide core was well screened. 
 Nevertheless, as shown by a decimation procedure~\cite{klotsa05}, 
 disorder in 
the backbone sites can induce local fluctuations of the onsite energies 
on the base pairs (gating effect). Uniform disorder (where the onsite 
energies of the backbones continuously  vary over an interval $[-W,W]$, 
$W$ being the
disorder strength) 
is shown to reduce continuously the localization length, as expected. 
For binary disorder (onsite 
energies  take only two posible values $\pm W/2$ ), 
as it may arise by the binding of counterions to the backbone sites, the 
situation is similar up to some
 critical disorder strength $W_c$. However, farther 
increase of $W$ leads to an unexpected behavior: the localization 
length on the electronic side bands is suppressed but a new band 
around the mid-gap with {\em increasing} localization length shows up.  
Thus, disorder-induced delocalization of the electronic 
states is observed in some energy window.  This result, obtained within 
a simple model, may be supported by   
 first principle calculations~\cite{endres05} which clearly show that 
 the environment can introduce additional states in the molecular 
band gap.

Most of the foregoing investigations considered onsite disorder, only. The
influence of off-diagonal short range correlations was investigated by Zhang and
Ulloa~
\cite{ulloa04} in $\lambda$-DNA. They showed that this kind of disorder can definitely 
lead to the emergence of 
conduction chanels in finite systems. For some special ratios of 
the nearest-neighbor hopping amplitudes, there may even exist  extended 
states in the thermodynamic limit. As a consequence, the authors 
suggested that  $\lambda$-DNA 
may show  a finite current at low voltages.

Caetano and Schulz~\cite{caetano04} investigated a double-strand model 
with uncorrelated disorder along the single strand, 
but taking into account the binding specificity of the four bases when 
considering the complementary strand (A-T and G-C).
Participation ratios $P(E)$ were computed, which give a measure of the degree 
of localization
of electronic states. $P(E)$ is e.~g. almost zero for localised states in the
 thermodynamic limit. The results suggest 
that inter-strand correlations may give rise to bands of delocalised 
states, with a participation ratio that does not appreciably 
decay with increasing length.

\section{Dynamical Disorder}

In the previous  section, we presented several studies related to the 
influence of
 static disorder on the charge transport 
properties of different DNA oligomers. Here, we address a second 
aspect of 
high relevance, namely the impact of 
dynamical disorder related to structural fluctuations, on charge
 propagation. Considering the relative flexibility of DNA, one may expect 
 that vibrational modes may have a strong influence on the charge motion 
 via a modification of electronic couplings.
 
  The considerably small decay rates found in electron-transfer 
experiments \cite{murphy93} have led to 
the proposal 
that, besides unistep superexchange mechanisms,  phonon-assisted 
 hole hopping might also be of importance \cite{jortner98}. 
The hole can occupy a specific  molecular orbital, localised on a 
given molecular site; it can also, however,  
extend over several molecular sites and build a  polaron, which is
 basically a lattice deformation acompanying a propagating 
charge. It results from the energetic interplay of two tendencies: 
the tendence to delocalise the charge, thus gaining kinetic energy, and  
 the tendence to localise it with a consequent gain
in elastic energy. The softness of the DNA molecule and the 
existence of modes that can appreciably affect the inter-base 
electronic coupling (like twisting modes or H-stretching bonds), 
makes this suggestion very attractive~\cite{hender99,conwell00}. 
Conwell and Rakhmanova~\cite{conwell00} investigated this issue using 
the Su-Heeger-Schrieffer (SSH) model, which is known 
to entail a rich nonlinear physics and that it has extensively been 
aplied to study polaron formation in conducting polymers.  
The SSH model deals classically with the lattice degrees of freedom 
while treating the electrons quantum mechanically. The 
calculations showed that a polaron may be built and be robust within 
a wide range of  model parameters. The influence 
of random base sequences was apparently  not strong enough to destroy it. 
Thus, polaron drifting may constitute a possible 
transport mechanism in DNA oligomers. 

The potential for the lattice 
displacements was assumed in Refs.~\cite{hender99,conwell00} to be
harmonic. Inter-strand modes like H-bond stretching are however expected to
 be strongly anharmonic; H-bond fluctuations can induce local
breaking of the double-strand and have thus been investigated in relation 
to the DNA denaturation 
problem ~\cite{peyr89}. To investigate this effect, Komineas et 
al.~\cite{komineas02} studied a model with strong 
anharmonic potentials and local coupling of the lattice to the charge 
density.  The strong nonlinearity of the problem led 
to a {\em dynamical} opening of bubbles with different sizes that may 
eventually trap the polaron and thus considerably affect this charge transport
channel. 

Zhang and Ulloa~\cite{ulloa02,zhang04} studied a simple  model 
that describes the coupling of torsional excitations (twistons) 
in DNA to propagating
charges and showed that this interaction  leads to polaron formation. 
Twistons  modify the 
inter-base electronic coupling, though this effect 
is apparently less strong than e.~g., in the Holstein model~\cite{holstein59},
 because of the  strong  nonlinearity of the twistons restoring forces as well 
 as of the twiston-electron coupling.  
 For small restoring forces of the twisting modes and in the non-adiabatic 
 limit ("spring constant" much bigger than electronic 
coupling), the inter-base  coupling is maximally perturbed and 
an algebraic band reduction is found, weaker than the exponential 
dependence known from the Holstein model.  Thus, it may be expected 
that the polaron will have a higher mobility along the chain.

The observation of two quite different time scales ($5\,ps$ and $75\,ps$) in the 
decay rates  of electron transfer processes in DNA, as measured by 
femtosecond spectroscopy~\cite{wan99}, was the main motivation 
of Bruinsma {\em et al.}~\cite{bruinsma00} to investigate 
the coupling of the electronic system to collective modes of the DNA cage. 
For this, they considered a tight-binding model of electrons 
interacting with two modes:
a twisting mode which mainly couples to the inter-base 
$\pi$-orbital matrix elements, 
and a linear displacement coupling to the onsite 
energies of the radical and acting as a local gating of the latter.  
 In the strong-coupling, high-temperature limit, 
 the hopping matrix elements can be 
treated perturbatively and build the lowest energy scale.  Transport 
has thus hopping-like character.
 In analogy with electron transfer 
theories, the authors provide a picture where there are basically  
two reaction coordinates related to the above mentioned
linear and angular modes. The strong thermal fluctuations associated 
with the twisting motion are shown to introduce 
two time scales for electron transfer that can be roughly related to optimal (short) 
and non-optimal (long) relative orientation of neighboring 
base pairs.

 In several papers,  Hennig {\em et al.}~\cite{hennig01,hennig04,yamada04}
 formulated a model Hamiltonian where only the relative 
transverse vibrations of bases belonging to the same 
pair are included. Their calculations showed the formation of stable 
polarons. 
Moreover, the authors suggested that poly(GC) should be more
effetive in supporting polaron-mediated charge transport than poly(AT), 
since for the latter the electron-lattice coupling was found to be about 
one order of magnitude smaller. 
Though the authors remarked that no appreciable 
coupling to twisting distortions  was found by their semiempirical 
quantum chemical calculations, this  issue requires further investigation in
view of the previously presented
results~\cite{ulloa02,zhang04,bruinsma00}. 
 Disorder did not apparently have a very dramatic 
 influence in this model; the localization length only changed 
 quantitatively as
 a function of the disorder strength~\cite{yamada04}.
 
 Asai~\cite{asai03} proposed a small 
 polaron model to describe the experimental findings 
 of Yoo {\em et al.}~\cite{yoo01} concerning the temperature dependence
 of the electric current  and of the linear conductance. Basically, he 
 assumed that in poly(GC) completely incoherent polaron hopping 
 dominates while in 
 poly(AT) quasi-coherent hopping, i.~e.~with total phonon number conservation,
  is 
 more important. As a result, the temperature dependence of the above quantites
 in both molecules
 is considerably different. 

Complementary to the foregoing research which mainly addressed individual 
vibrational  modes of the DNA cage, other studies have
focused on  the influence of environmental effects.
Basko and
 Conwell~\cite{basko02} used a semiclassical model to describe the 
 interaction of an
injected hole in DNA which is placed in a polar solvent. Their basic 
conclusions  pointed out  that the main contribution 
was given by the interaction 
with  water molecules and not with counterions; further, polaron formation 
was not hindered by the charge-solvent coupling, the interaction
 rather increased the binding energy (self-localization) of the polaron 
 by around half an eV, which is much larger than relevant temperature scales.
 Li and Yan~\cite{li01} as well as  Zhang {\em et al.}~\cite{zhang02} 
investigated the role of dephasing reservoirs 
 in the spirit of the B{\"u}ttiker-D'Amato-Pastawski model~\cite{Bu88,pasta90}. 
 Zhang {\em et al.}  showed that a change in the length scaling of the
 conductance can be induced by the dephasing reservoirs as a result of 
 incoherent phonon-mediated transport, a result known from electron transfer
 theories~\cite{segal00}.  In a
 similar  way, Feng and Xiong~\cite{feng02} considered gap-opening as resulting 
 from the coupling to a set 
 of two-level systems which simulate low-lying states of the bosonic bath. 
 Gutierrez {\em et al.}~\cite{gmc05a,gmc05b} have discussed electron transport
 in a 
 ``broken''-ladder model in presence of a strong dissipative environment
 simulated by a bosonic bath. It was found that the environment can induce 
 virtual polaronic  states inside the molecular band gap and thus lead to a change 
 in the low-energy transport properties of the system. Especially, the $I$-$V$ 
 curves become metallic-like at low voltages as a result of phonon-assisted 
 hopping. We note that these latter results are quite similar to that found in {\em ab
 initio} calculations, showing that water states can appear inbetween the 
 $\pi-\pi^{*}$ gap~\cite{endres04}, thus effectively introducing shallow states similar to those in doped bulk semiconductors. These states may support activated hopping at high temperatures.

We finally mention that the role  of nonlinear excitations 
(solitons, breathers) 
in the process of denaturation of DNA double 
strands~\cite{peyr89,xiao87,yaku02} and in the transmission of  
``chemical''
information between remote DNA segments~\cite{hermon98} 
have been early addressed in the literature.  Since these approaches 
are not directly connected with the issue of charge transport 
in  DNA wires between electrodes, we do not 
go into further details. They may however open a new interesting mechanism 
for transport and deserve a more careful investigation.

\section{Conclusions}
Though big progress has been achieved in the past decade to clarify the 
relevant transport mechanisms in DNA oligomers, a coherent, unifying picture is still 
lacking. The experimental difficulties to give reliable transport
characteristics of this molecule make the formulation of model Hamiltonians
quite challenging. The theoretical research presented in this chapter 
shows that charge transport in DNA is considerably influenced by both 
static and dynamical disorder. Long-range correlated disorder 
can play a role in increasing the localization length beyond the relevant 
molecular length scales addressed in experiments, thus making DNA to effectively 
appear as 
 a conductor. This effect may be supported or counteracted by thermal
 fluctuations arising from internal (vibrations) or external (solvent) modes and 
 leading to increased charge localization or to  incoherent 
 transport. 
 
 The presented models only focus on  the equilibrium, low-bias limit of
 transport. However, real transport experiments probe the molecules at finite voltages
 and hence, non-equilibrium effects have to be also considered. This makes 
 of course the mathematical treatment as well as the physical 
 interpretation  more involved. Considerable efforts 
 to deal with this issue  have been 
 made in the last times~\cite{ventra05,pecchia04,galperin05}; 
 to address them  goes however beyond the scope of this chapter. 

\section{Acknowledgments}
The authors thank R. Bulla, A. Nitzan, R. R{\"o}mer and S. Roche  for 
useful suggestions and discussion. 
This work has been supported by the Volkswagen Foundation and by the EU under
contract IST-2001-38951.
%
 
%
%
%
%
%

\end{document}